\documentclass[11pt,a4paper,english]{article}
\usepackage{lmodern}

\usepackage[T1]{fontenc}
\usepackage[latin9]{inputenc}
\usepackage{amsmath}
\usepackage{amssymb}

\makeatletter

\pdfpageheight\paperheight
\pdfpagewidth\paperwidth

\usepackage{jcappub}
\numberwithin{equation}{section}

\pdfoutput=1
\renewcommand\[{\begin{equation}}
\renewcommand\]{\end{equation}}

\makeatother

\usepackage{babel}
\begin{document}

\title{Superconducting Dark Matter }

\author{Alexander Vikman}

\affiliation{CEICO-Central European Institute for Cosmology and Fundamental Physics, }

\affiliation{Institute of Physics of the Czech Academy of Sciences, \\
Na Slovance 2, 182 21 Prague 8, Czech Republic\\
}

\emailAdd{vikman@fzu.cz}

\abstract{We extend recently proposed mimetic dark matter (DM) and dust of
dark energy (DE) models to a U(1)-charged gauged scalar field. This
extension yields a mass to the photon via the Higgs mechanism. The
square of this photon mass is proportional to the DM energy density.
Thus dense DM environments can screen the dark vector force. There
is a substantial freedom in this gauged extension of the irrotational
fluid-like DM. This freedom enables one to model the classical London
equation of superconductivity or to obtain the photon mass which remains
constant during the matter-dominated époque. }
\maketitle

\section{Introduction}

Observations show\footnote{For review see e.g. \cite{Durrer:2013pga}\cite{Grasso:2000wj}}
\cite{Neronov:1900zz} that magnetic fields are present not only in
galaxies, but also in intergalactic space on cosmological scales.
The structure of these fields and the history of their evolution is
rather complicated and heavily influenced by different astrophysical
properties. The origin of these large-scale cosmological magnetic
fields remains as unclear as the origin of dark matter. It would be
great if inflation could produce proper seeds, as it was first suggested
in \cite{Turner:1987bw}, but it seems that it is not an easy task
see e.g. \cite{Durrer:2013pga}\cite{Demozzi:2009fu}. 

Classical electrodynamics is conformally invariant, therefore photons
cannot not be produced in evolving cosmological backgrounds, as the
conformal vacuum is preserved in this case \cite{Parker:1968mv}.
Thus the conformal invariance of the electromagnetism must be broken
to produce long rage magnetic fields as a result of inflation. The
simplest way to break conformal invariance is to introduce the mass
to the photon. It is well known, that superconducting media forces
the photon to become massive. This is very similar to the Higgs effect.
The main difference is that Higgs effect yields the mass which is
constant and does not depend on the environment, whereas the London
equation introduces the mass proportional to the square root of the
charge carrier density. 

On the other hand, it is probably too naive and definitely premature
to claim that the dark sector of the universe has a simple structure
and cannot contain dark $U\left(1\right)$ field. Moreover, gauging
the only anomaly-free global symmetry of the Standard Model - the
difference between baryon number B and lepton number L could give
rise to another $U\left(1\right)$ field. It is interesting whether
these dark long range forces can be screened and avoid observations
so far in this way. Making these dark photons to leave in a superconductor
would enable the desired screening. 

It seems that it is interesting to equip the dark sector with superconducting
properties. In this paper we present a class of models which make
a fluid-like DM to behave as a superconductor. In particular, we managed
to obtain the London equation with the square of the photon mass proportional
to the DM density. 

\section{Model }

Standard irrotational dust can be described by the action 
\begin{equation}
S=\int d^{4}x\sqrt{-g}\,\frac{\rho}{2}\,\left(\frac{\left(\partial\varphi\right)^{2}}{M^{4}}-1\right)\,,\label{eq:dust_action}
\end{equation}
where $\rho$ is a Lagrange multiplier playing the role of the energy
density while the \emph{real} scalar field $\varphi$ is the velocity
potential\footnote{We use: the standard notation $\sqrt{-g}\equiv\sqrt{-\text{det}g_{\mu\nu}}$
where $g_{\mu\nu}$ is the metric, the signature convention $\left(+,-,-,-\right)$,
and the units $c=\hbar=1$, $M_{\text{Pl}}=\left(8\pi G_{\text{N}}\right)^{-1/2}=1$.
However, sometimes for convenience we explicitly write $M_{\text{Pl}}$. }. The mass scale $M$ is introduced to keep the canonical dimensions
of the noncanonical scalar field $\varphi$. This action (\ref{eq:dust_action})
also naturally appears \cite{Golovnev:2013jxa,Barvinsky:2013mea,Hammer:2015pcx}
in the so-called \emph{mimetic gravity }\cite{Chamseddine:2013kea,Chamseddine:2014vna}\emph{.
}For later it is useful to note that the mass scale $M$ can be assumed
to be a function of the field $\varphi$, so that one can allow for
a function 
\[
M\left(\varphi\right)=\mu f\left(\varphi/\mu\right)\,,
\]
where $f$ is a free function and $\mu$ is a constant mass scale.
This extension does not change the dynamics, as the following field-redefinition\footnote{Similar redefinition, along with an obvious redefinition of the Lagrange
multiplier reduces the constraint in \cite{Lim:2010yk} to the form
(\ref{eq:dust_action}) with a constant mass-scale $\mu$. } 
\begin{equation}
d\phi=\frac{d\varphi}{f^{2}\left(\varphi/\mu\right)}\,,\label{eq:redefinition}
\end{equation}
maps this theory back to (\ref{eq:dust_action}) with $\varphi\rightarrow\phi$
and $M\left(\varphi\right)\rightarrow\mu$. We will later use this
fact. 

In this paper we generalize the theory (\ref{eq:dust_action}) to
a \emph{complex} scalar field $\varphi$ interacting with a $U\left(1\right)$
gauge field $A_{\mu}$ in the standard covariant way 

\begin{equation}
S_{\varphi}=\int d^{4}x\sqrt{-g}\,\frac{\rho}{2}\,\left(\frac{\left|D_{\mu}\varphi\right|^{2}}{M^{4}}-1\right)\,,\label{eq:charged_dust}
\end{equation}
where the $U\left(1\right)$- covariant derivative extends the usual
Levi-Civita connection $\nabla_{\mu}$ as 
\begin{equation}
D_{\mu}=\nabla{}_{\mu}+iqA_{\mu}\,,\label{eq:covariant_derivative}
\end{equation}
so that $q$ is the coupling constant of $\varphi$ to $A_{\mu}$.
This action can be obtained from the \emph{mimetic} \emph{substitution}
into the standard action for gravity $g_{\mu\nu}$ and matter
\begin{equation}
g_{\mu\nu}=\frac{h^{\alpha\beta}D_{\alpha}\varphi\overline{D_{\beta}\varphi}}{M^{4}}\cdot h_{\mu\nu}\,,\label{eq:Mimetic_substitution}
\end{equation}
which is\emph{ }a particular Weyl (conformal) transformation of the
auxiliary metric $h_{\mu\nu}$. This substitution results in a degenerate
higher derivative and Weyl-invariant and $U\left(1\right)$ gauged-scalar
($\varphi$)-tensor ($h_{\alpha\beta}$)-vector $\left(A_{\mu}\right)$
theory. 

We assume that the dynamics of the free gauge field is given by the
standard action 
\begin{equation}
S_{a}=-\frac{1}{4}\int d^{4}x\sqrt{-g}\,F_{\mu\nu}F^{\mu\nu}\,,\label{eq:Maxwell_action}
\end{equation}
with the field strength $F_{\mu\nu}=\partial_{\mu}A_{\nu}-\partial_{\nu}A_{\mu}$.
Gravity is minimally coupled to matter and to the scalar field and
is described by the Einstein-Hilbert action. Both actions (\ref{eq:charged_dust})
and (\ref{eq:Maxwell_action}) are invariant under the gauge transformation
\begin{align*}
\varphi\rightarrow e^{-iq\lambda}\varphi\,,\qquad & A_{\mu}\rightarrow A_{\mu}+\partial_{\mu}\lambda\,.
\end{align*}

It is important to stress that, contrary to the original irrotational
dust described by the action (\ref{eq:dust_action}), the shift-symmetry
in field space $\varphi\rightarrow\varphi+c$, is broken in the presence
of the gauge field $A_{\mu}$. Hence it is not forbidden by any symmetry
to introduce a potential, $V\left(\left|\varphi\right|\right)$, or
/ and to promote the scale $M$ to be a function of the field: $M\left(\left|\varphi\right|\right)$\footnote{One could also promote $q\rightarrow q\left(\left|\varphi\right|\right)$,
but we will not pursue this option in this paper.}. The shift-symmetry breaking makes the dynamics similar to the \emph{Dust
of Dark Energy }\cite{Lim:2010yk} and allow to unify DM and DE.\emph{
}Moreover, this construction can motivate other shift-symmetry breaking
terms used in \cite{Chamseddine:2013kea,Lim:2010yk,Chamseddine:2014vna,Mirzagholi:2014ifa}.
It is crucial to mention that some shift-symmetry breaking is needed
for these models to survive inflation. Let us leave the case with
potential for a future work, but allow for a nontrivial mass scale
in (\ref{eq:charged_dust})
\begin{equation}
M\left(\left|\varphi\right|\right)=\mu f\left(\left|\varphi\right|/\mu\right)\,.\label{eq:mass_scale}
\end{equation}
The energy-momentum tensor (EMT) of the scalar sector is 
\begin{equation}
T_{\mu\nu}=\frac{2}{\sqrt{-g}}\,\frac{\delta S_{\varphi}}{\delta g^{\mu\nu}}=\frac{\rho}{2M^{4}}\left(D_{\mu}\varphi\overline{D_{\nu}\varphi}+\overline{D_{\mu}\varphi}D_{\nu}\varphi\right)\,.\label{eq:EMT_No_Potential}
\end{equation}
Due to the interaction with the vector field $A_{\mu}$ this EMT is
not conserved. The $U\left(1\right)$-conserved current sourcing the
Maxwell equations 
\begin{equation}
\nabla_{\mu}F^{\mu\nu}=J^{\nu}\,,\label{eq:Maxwell}
\end{equation}
is given by 
\begin{equation}
J_{\mu}=-\frac{1}{\sqrt{-g}}\,\frac{\delta S_{\varphi}}{\delta A^{\mu}}=\frac{iq\rho}{2M^{4}}\,\left(\bar{\varphi}D_{\mu}\varphi-\varphi\overline{D_{\mu}\varphi}\right)\,.\label{eq:conserved_current}
\end{equation}
Other equations of motion are: the constraint 
\begin{equation}
\left|D_{\mu}\varphi\right|^{2}=M^{4}\,,\label{eq:Constraint}
\end{equation}
and equation of motion for the scalar field
\begin{equation}
D_{\mu}\left(\frac{\rho}{M^{4}}\,D^{\mu}\varphi\right)=0\,,\label{eq:Phi_eom}
\end{equation}
along with its complex conjugated. Because of the constraint (\ref{eq:Constraint}),
either the field-derivatives $\partial_{\mu}\varphi$ or the gauge
field $A_{\mu}$ (or both) are never vanishing. Moreover, either the
absolute value of the field $\left|\varphi\right|$ or its derivatives
are always not vanishing as well. 

For the later it is convenient to use the polar decomposition of the
scalar field 
\begin{equation}
\varphi=\chi\,e^{iq\theta}\,,\label{eq:polar_decomposition}
\end{equation}
which is a nonsingular field-reparametrization provided $\chi\neq0$.
Using this parametrization we obtain 
\begin{equation}
D_{\mu}\varphi=e^{iq\theta}\left(\partial_{\mu}\chi+iq\chi\left(\partial_{\mu}\theta+A_{\mu}\right)\right)\,,\label{eq:Covariant_derivative_in_polar}
\end{equation}
so that it is natural to introduce a \emph{gauge-invariant} vector
field 
\begin{equation}
G_{\mu}\equiv A_{\mu}+\partial_{\mu}\theta\,.\label{eq:Gauge_inv_vector}
\end{equation}
In terms of this new gauge-invariant dynamical variables $\chi$ and
$G_{\mu}$ we have for the constraint equation (\ref{eq:Constraint})
\begin{equation}
\left|D_{\mu}\varphi\right|^{2}=\left(\partial_{\mu}\chi\right)^{2}+q^{2}\chi^{2}G^{\mu}G_{\mu}=M^{4}\,.\label{eq:Kinetic_Term}
\end{equation}
Hence by virtue of the polar decomposition (\ref{eq:polar_decomposition})
the new gauge-invariant vector field $G_{\mu}$ becomes massive with
the mass given by 
\begin{equation}
m_{G}^{2}=\frac{\rho}{M^{4}}\,q^{2}\chi^{2}\,.\label{eq:Mass_general_formula}
\end{equation}
The best way to detect the mass of the vector field $G_{\mu}$ is
to notice that the conserved current (\ref{eq:conserved_current})
becomes 
\begin{equation}
J_{\mu}=-m_{G}^{2}\,G_{\mu}\,,\label{eq:Current_higgs}
\end{equation}
and introduces a mass term (\ref{eq:Mass_general_formula}) in the
Maxwell equations. One can expect this mass, $m_{G}$, to be generically
time-dependent. The other decisive difference from the standard canonical
abelian Higgs mechanism is the prefactor $\rho/M^{4}$. Here the dependence
of the mass-scale on the field absolute value $\chi$ becomes crucial.
For example the mass scale 
\begin{equation}
M\left(\chi\right)=\sqrt{\mu\,\chi}\,,\qquad\text{results in }\qquad m_{G}^{2}=q^{2}\,\frac{\rho}{\mu^{2}}\,,\label{eq:mass_eq_density}
\end{equation}
so that the mass of the vector field does not depend on the scalar
field $\chi$ at all. In this case the current is 
\begin{equation}
\mathbf{J}=-q^{2}\,\frac{\rho}{\mu^{2}}\mathbf{G}\text{ ,}\label{eq:London_equatioin}
\end{equation}
which gives the classical London equation in the superconductivity
theory. Hence our dark matter is a \emph{superconductor}. In fact
our system with mass scale $M\left(\chi\right)$ above (\ref{eq:mass_eq_density})
exactly realizes the London equation in a manifestly covariant and
gauge invariant manner. The mass scale $\mu$ can be associated with
the mass of the charged particle. It is important to note that this
way of choosing the mass scale $M\left(\chi\right)$ is rather nontrivial
and forces the Lagrangian to be not an analytic function of fields
around zero. 

The EMT (\ref{eq:EMT_No_Potential}) reads in the new variables as
\begin{equation}
T_{\mu\nu}=\frac{\rho}{M^{4}}\left(\chi_{,\mu}\chi_{,\nu}+q^{2}\chi^{2}G_{\mu}G_{\nu}\right)\,.\label{eq:EMT_higgs}
\end{equation}
The current conservation $\nabla_{\mu}J^{\mu}=0$ yields 
\begin{equation}
\nabla_{\mu}\left(\frac{\chi^{2}\rho}{M^{4}}\,G^{\mu}\right)=0\,.\label{eq:current_conservation}
\end{equation}
In particular, in cosmology this implies that the charge density redshifts
as 
\begin{equation}
\frac{\chi^{2}\rho}{M^{4}}\,G^{0}=\frac{c}{a^{3}}\,.\label{eq:chargedensity_redshift}
\end{equation}
For timelike gradients of the radial field $\chi$ one can introduce
normalized vectors 
\begin{equation}
u_{\mu}=\frac{\chi_{,\mu}}{\sqrt{\left(\partial\chi\right)^{2}}}\,,\label{eq:velocity}
\end{equation}
which one can use as a local rest frame so that 
\begin{equation}
T_{\mu\nu}=\mathcal{E}u_{\mu}u_{\nu}+m_{G}^{2}\,G_{\mu}G_{\nu}\,,\label{eq:EMT_u_G}
\end{equation}
where 
\[
\mathcal{E}=\frac{\rho\left(\partial_{\mu}\chi\right)^{2}}{M^{4}}\,.
\]
The total conserved EMT\footnote{This total conserved EMT differs from the EMT for the Proca vector
field and dust, by the absence of the $-\frac{1}{2}g_{\mu\nu}m_{G}^{2}\,G_{\alpha}G^{\alpha}$
term. This term is absent due to the constraint equation (\ref{eq:Constraint})
which we used. } is 
\[
T_{\mu\nu}^{+}=\mathcal{E}u_{\mu}u_{\nu}+F_{\mu\lambda}F_{\;\;\;\nu}^{\lambda}+\frac{1}{4}g_{\mu\nu}\,F_{\alpha\beta}F^{\alpha\beta}+m_{G}^{2}\,G_{\mu}G_{\nu}\,,
\]
with the field strength tensor $F_{\mu\nu}=\partial_{\mu}G_{\nu}-\partial_{\nu}G_{\mu}$.

From the real part of the equation (\ref{eq:Phi_eom}) we obtain\footnote{Imaginary part of this equation gives the current conservation (\ref{eq:current_conservation}).}
\[
\nabla_{\mu}\left(\rho\nabla^{\mu}\chi\right)=q^{2}\rho\chi G^{\mu}G_{\mu}\,.
\]
This equation is homogeneous in $\rho$ and admits the trivial vacuum
solution $\rho=0$. 

The field $\rho$ moves almost as the dust energy density except of
a small $\mathcal{O}\left(q^{2}\right)$ external force on right hand
side of
\[
\dot{\rho}+\Theta\rho=q^{2}\rho\,\left(\frac{2M^{2}\chi G^{2}+u^{\lambda}\nabla_{\lambda}\left(\chi^{2}G^{2}\right)}{2M^{4}}+\mathcal{O}\left(q^{2}\right)\right)\,,
\]
where $\dot{\rho}=u^{\lambda}\nabla_{\lambda}\rho$ and $\Theta=\nabla_{\mu}u^{\mu}$
is the expansion of the $u^{\mu}$ congruence. In cosmology, the vector
field can be only present on the level of perturbations. Hence 
\[
\dot{\rho}+\Theta\rho\simeq0\,,
\]
up to small perturbations terms multiplied with a small $q^{2}$.
The congruence $u^{\mu}$ is almost geodesic as 
\[
a_{\mu}=u^{\lambda}\nabla_{\lambda}u_{\mu}=\nabla_{\mu}^{\perp}\ln\left(\partial\chi\right)^{2}\simeq-\frac{q^{2}}{M^{4}}\nabla_{\mu}^{\perp}\left(\chi^{2}G^{2}\right)+\mathcal{O}\left(q^{4}\right)\,,
\]
where we used projector $\bot_{\mu\nu}=g_{\mu\nu}-u_{\mu}u_{\nu}\,,$
to introduce spatial derivative $\nabla_{\mu}^{\perp}=\bot_{\mu}^{\nu}\nabla_{\nu}\text{ .}$
Hence in the limit 
\begin{equation}
q^{2}\chi^{2}\left|G^{\mu}G_{\mu}\right|\ll M^{4}\,,\label{eq:fluid_condition}
\end{equation}
 the filed $\chi$ plays the role of the velocity potential for dust
while $\rho$ is its energy density. In this case, on the background
level in cosmology $\left(\partial_{\mu}\chi\right)^{2}\simeq M^{4},$
so that $\mathcal{E}\simeq\rho$ . Hence this can be called ``Higgs
dust'' or superconducting DM as because of the effective photon mass
and the London equation (\ref{eq:London_equatioin}). 

For $M=const$ 
\begin{equation}
\chi\simeq M^{2}t\,,\label{eq:Chi_t}
\end{equation}
while for more general mass-scales (\ref{eq:mass_scale}) one can
use the redefinition (\ref{eq:redefinition}) to obtain
\begin{equation}
\int\frac{d\chi}{f^{2}\left(\chi/\mu\right)}=\mu^{2}t\,,\label{eq:CHI_of_time}
\end{equation}
which gives an implicit function $\chi\left(t\right)$. 

In this regime, when one can neglect the energy density stored in
the massive term of the vector field, the energy density of the dust
redshifts in the standard way 
\[
\rho\propto a^{-3}\,,
\]
while the freedom in the choice of the mass-scale function $f\left(\chi/\mu\right)$
allows one to arrange for practically \emph{arbitrary} evolution of
$m_{G}\left(t\right)$. In particular, the mass redshifts slower than
the dust energy density provided 
\[
\frac{d}{dt}\left(\frac{\chi}{M^{2}\left(\chi\right)}\right)>0\,.
\]
Which gives 
\[
\frac{\dot{\chi}}{M^{2}}\left(1-\frac{2\chi M'}{M}\right)=1-\frac{2\chi M'}{M}>0\,,
\]
or 

\[
\frac{d\ln M}{d\ln\chi}<\frac{1}{2}\,.
\]

The equality in the last expression would yield the mass scale (\ref{eq:mass_eq_density}).
For $M=const$ one can neglect the energy density stored in the massive
term of the vector field provided 
\begin{equation}
q^{2}t^{2}\left|G^{\mu}G_{\mu}\right|\ll1\,.\label{eq:condition_applicability_for_dust}
\end{equation}
If we estimate the relevant time to be the cosmological time 
\[
t\simeq H^{-1}\,,
\]
we obtain
\[
q^{2}\left|G^{\mu}G_{\mu}\right|\ll H^{2}\,.
\]
Let's first assume that there is only magnetic field $B_{\ell}$ on
characteristic scale $\ell$. Then $G^{\mu}G_{\mu}=-\mathbf{G}^{2}$
so that $\left(\partial\chi\right)^{2}>0$ and $\chi$ can be still
used as a potential for some four-velocity. The condition that magnetic
field corrections do not spoil the dust-like behavior of the in the
Higgs phase implies that the charge is bounded by 
\begin{equation}
q\ll\frac{H}{B_{\ell}\ell}\,.\label{eq:small_charge_magnetic}
\end{equation}
If the system violates this bound it i) either cannot be in the Higgs
phase and form some magnetic tubes / strings or ii) the system does
not behave like dust and has a strong influence of magnetic field
on the motion. 

In this paper we consider a general $U\left(1\right)$ field. However,
we can try to apply this to the standard electromagnetic interactions
in the Standard Model (SM). There are seem to be magnetic fields on
large inter-cluster scales \cite{Neronov:1900zz} $B\sim10^{-16}\,\text{G}\sim10^{-35}\,\text{GeV}^{2}\sim10^{-73}$
on scales of $\ell\sim1\,\text{Mpc}\sim3\times10^{24}\text{\text{m}}\sim10^{57}$
so that 
\[
q\ll\frac{10^{-62}}{10^{-73}10^{57}}\simeq10^{-46}
\]

Further for galactic magnetic field one can assume $B\sim\mu\text{G}\sim10^{-25}\,\text{GeV}^{2}\sim10^{-63}$,
$\ell\sim1\,\text{kpc}\sim3\times10^{21}\text{\text{m}}\sim10^{54}$
and $H\sim10^{-61}$ so that 

\[
q\ll\frac{10^{-61}}{10^{-63}10^{54}}\simeq10^{-52}\,.
\]
On the other hand for the case reproducing the London equation (\ref{eq:mass_eq_density})
we have 
\[
q^{2}\chi^{2}\left|G^{\mu}G_{\mu}\right|\ll\mu^{2}\chi^{2}\text{ ,}
\]
so that 
\begin{equation}
G\ll\frac{\mu}{q}\text{ .}\label{eq:Bound_on_G_London}
\end{equation}
For a magnetic field on scale $\ell$ this bound gives 
\[
B_{\ell}\ell\ll\frac{\mu}{q}\text{ ,}
\]
which for scales $\ell\simeq\mu^{-1}$ for the electric field corresponds
to the Schwinger limit. If we apply this boldly to the SM electrodynamics
we get 
\begin{equation}
q\ll\frac{\mu}{B_{\ell}\ell}\simeq\left(\frac{\mu}{M_{\text{Pl}}}\right)\times10^{9}\text{ .}\label{eq:bound_on_q_galaxy}
\end{equation}

\section{``Photon'' mass in cosmology, theoretical and observational constraints }

In case of the Higgs dust with (\ref{eq:mass_eq_density}) reproducing
the London equation, the mass of the photon always redshifts as $a^{-3}$.
On the other hand a naively more natural choice $M=const$ gives a
more interesting evolution $m_{G}\left(t\right)$. Below we consider
different cases. 

\subsection{Matter domination }

In the universe completely dominated by this Higgs dust we have 
\[
\rho=3M_{\text{Pl}}^{2}H^{2}=\frac{4}{3}M_{\text{Pl}}^{2}\cdot\frac{1}{t^{2}}\,,
\]
so that the mass of the $U\left(1\right)$-vector field (\ref{eq:Mass_general_formula})
is \emph{time-independent} for the constant mass scale $M$
\begin{equation}
m_{G}=\frac{2}{\sqrt{3}}q\,M_{\text{Pl}}\,.\label{eq:Mass_G_constant}
\end{equation}
In the DM-dominated and spatially-flat universe where only $\Omega_{H}$
fraction of energy is in the Higgs dust we have 
\[
\rho=3M_{\text{Pl}}^{2}H^{2}\Omega_{H}\,,
\]
so that the photon mass is again \emph{time-independent} 
\begin{equation}
m_{G}=q\sqrt{\frac{4\Omega_{H}}{3}}\,M_{\text{Pl}}\,,\label{eq:mass_matter_domination}
\end{equation}
where $\Omega_{H}$ remains constant. To develop intuition we can
boldly apply this construction to the electrodynamics from the SM.
Then we can use the astrophysical constraint, (see discussion in \cite{Adelberger:2003qx})
on the photon mass 
\begin{equation}
m_{\gamma}^{a}\lesssim10^{-26}\,\text{eV}\,,\label{eq:constraint_on_mass}
\end{equation}
and obtain that 
\[
q\sqrt{\frac{\Omega_{H}}{6\pi}}\lesssim\frac{10^{-26}\,\text{eV}}{1.22\times10^{28}\,\text{eV}}\simeq10^{-54}\,.
\]
On the other hand, if our system works as a type II superconductor
where the coherence length is smaller than the penetration depth $m_{G}^{-1}$
this astrophysical bound is invalid \cite{Adelberger:2003qx}. Then
we need to use the much weaker laboratory Coulomb law bound 
\begin{equation}
m_{\gamma}^{l}\lesssim10^{-14}\,\text{eV}\,,\label{eq:Laboratory_bound}
\end{equation}
which still yields a tide constraint 
\[
q\sqrt{\frac{\Omega_{H}}{6\pi}}\lesssim\frac{10^{-14}\,\text{eV}}{1.22\times10^{28}\,\text{eV}}\simeq10^{-42}\text{ .}
\]
In any case, either the fraction of the Higgs dust (with constant
mass scale $M\left(\chi\right)=const$) in the energy budget is vanishingly
small or the coupling constant is extremely tiny.

It is interesting to mention that there is the so-called \emph{Gravity
as the Weakest Force} (GWF) conjecture \cite{ArkaniHamed:2006dz}
which states that the cutoff in gauge theories is 
\[
\Lambda\simeq q\,M_{\text{Pl}}\,.
\]
If we assume that this condition is also applicable to the environmental
masses which are not fundamental we have $m_{G}\lesssim q\,M_{\text{Pl}}$
and therefore 
\[
\Omega_{H}\lesssim\frac{3}{4}\,.
\]
In this case the GWF conjecture (if applicable) forces the Higgs dust
with the $M=const$ to be a heavily subdominant component in DM budget. 

On the other hand, for the choice (\ref{eq:mass_eq_density}) reproducing
the London equation we have a time dependent mass 
\[
m_{G}\left(t\right)=\frac{q}{\mu}HM_{\text{Pl}}\sqrt{3\Omega_{H}}\lesssim q\,M_{\text{Pl}}\text{ . }
\]
 Hence the GWF provides a very weak bound from below on the mass of
the charge carrier 
\[
\mu\gtrsim H\sqrt{3\Omega_{H}}\text{ .}
\]
If we again boldly apply this construction for the usual electrodynamics
from the SM and use the constraint (\ref{eq:constraint_on_mass})
we obtain the bound on the mass of the electric charge carrier building
the superconductor 
\[
\mu\gtrsim q\sqrt{\frac{3\Omega_{H}}{8\pi}}\left(\frac{1.22\times10^{28}\,\text{eV}\text{ }10^{-33}\,\text{eV}}{10^{-26}\,\text{eV}}\right)\simeq q\sqrt{\frac{3\Omega_{H}}{8\pi}}\times10^{12}\,\text{GeV ,}
\]
which is stronger than (\ref{eq:bound_on_q_galaxy}) and can be clearly
satisfied by some heavy particle. If again our system works as a type
II superconductor, then we reserve to the weaker laboratory bound
(\ref{eq:Laboratory_bound}) which yields 
\[
\mu\gtrsim q\sqrt{\frac{3\Omega_{H}}{8\pi}}\,\,\text{GeV ,}
\]
which is weaker than (\ref{eq:bound_on_q_galaxy}). 

\subsection{Radiation domination}

During the radiation dominated époque $a\propto t^{1/2}$ so that
\[
\rho\propto a^{-3}\propto t^{-3/2}\,.
\]
so that for $M=const$ the mass of the gauge boson $m_{G}^{2}=\rho\,q^{2}t^{2}\,,$
grows with time to the maximal constant value (\ref{eq:mass_matter_domination})
as 
\[
m_{G}\propto t^{1/4}\,.
\]

\subsubsection{$\Lambda\text{CDM}$}

In the universe with a cosmological constant and DM the scale factor
evolves as, (see e.g. \cite{Mukhanov:2005sc})
\[
a\left(t\right)=a_{m}\left(\sinh\frac{3}{2}H_{\Lambda}t\right)^{2/3}\,,
\]
where $H_{\Lambda}=H_{0}\sqrt{1-\Omega_{m}}\,,$ and $\Omega_{m}=\Omega_{DM}+\Omega_{H}\,.$
After some straightforward manipulations one finds 
\[
m_{G}\left(t\right)=q\,M_{\text{Pl}}\sqrt{\frac{3\Omega_{H}}{\Omega_{m}}}\,\left[\frac{H_{0}\sqrt{1-\Omega_{m}}\,t}{\sinh\left(\frac{3}{2}\left(H_{0}\sqrt{1-\Omega_{m}}\right)t\right)}\right]\,.
\]
Hence the mass is decreasing in the late times when $\Lambda$ starts
to dominate. The maximal mass is 
\[
m_{G}\left(t\right)=q\,M_{\text{Pl}}\sqrt{\frac{4\Omega_{H}}{3\left(\Omega_{H}+\Omega_{DM}\right)}}\,.
\]

\subsection{Decay of perturbations }

The usual excitations of the Higgs field are unstable and decay. This
is also the case in our construction. On the perturbative level the
mass scale defines the coupling constants and decay rates for fluctuations
$\delta\rho$ and $\delta\chi$. Indeed, the cubic vertices in the
presence of a background $\chi$ are 
\begin{equation}
\left(\frac{q^{2}\chi^{2}}{M^{4}}\right)\delta\rho\,G^{\mu}G_{\mu}\,,\label{eq:Rho_GG}
\end{equation}
and 
\begin{equation}
q^{2}\rho\left(\frac{\chi^{2}}{M^{4}}\right)'\delta\chi\,G^{\mu}G_{\mu}\,,\label{eq:Chi_GG}
\end{equation}
where $\left(\,\right)'=d(\,)/d\chi$. It is interesting to note that
for (\ref{eq:mass_eq_density}) the last interaction vertex (\ref{eq:Chi_GG})
is vanishing. These couplings are time dependent. In particular, for
a constant $M$ both coupling constants are growing with time. 

However, both fields $\delta\chi$ and $\delta\rho$ are not independent
degrees of freedom and are not canonical wave-like fields with usual
propagators. Hence the standard field theory methods cannot be directly
applied here. Yet, it is clear that we are interested in rather small
coupling constants $q$ so that the decay time-scales can be longer
than the age of the universe. A proper estimation of the possible
instability rate due to the production of two massive ``photons''
lies beyond the scope of this work and will be addressed somewhere
else. 

\section{Discussion and Further Directions }

We presented a model of superconducting DM. In particular we managed
to derive the London equation. However, inclusion of a potential can
promote this to the model of superconducting dark energy. Or superconducting
unification of DM and DE. 

It is very interesting to work out perturbation theory in this models
and consider how this construction can influence the structure formation.
Of course it is important to work out observation constraints in more
details. Another issue is the quantization. Indeed, these models behave
like fluid-like dust and do not have propagating DM waves. In particular
it is important to see the limitations coming from potential strong
couplings. It is well known that this systems are plagued by caustics.
In the usual Higgs mechanism the theory restores the symmetry on short
scales and the mass disappears. Maybe this softer behavior on short
scales can help to solve these issues. 

It is also interesting to generalize this conformal \emph{gauged mimetic
substitution} (\ref{eq:Mimetic_substitution}) to promote the disformal
transformations \cite{Bekenstein:1992pj} in a similar way 
\[
g_{\mu\nu}=\omega\cdot h_{\mu\nu}+F\cdot D_{(\mu}\varphi\overline{D_{\nu)}\varphi}\,,
\]
where both functions $\omega$ and $F$ can depend on standard kinetic
term $h^{\alpha\beta}D_{\alpha}\varphi\overline{D_{\beta}\varphi}$
and on $\varphi\bar{\varphi}$. It was demonstrated before \cite{Deruelle:2014zza}
that singular disformal transformations do introduce new degrees of
freedom. 

Other possible extensions include direct couplings to curvature or
/ and incorporation of higher derivative operators which can give
a small sound speed to small fluctuations of $\varphi$, e.g. one
can add a term $\left|D_{\lambda}D^{\lambda}\varphi\right|^{2}$to
the Lagrangian. Further one can try to gauge other beyond Horndeski
theories. 

\acknowledgments It is a pleasure to thank Gia Dvali, Andrei Gruzinov,
Slava Mukhanov and Iggy Sawicki for useful discussions, encouragement
and criticism. This work was supported by the J. E. Purkyn\v{e} Fellowship
of the Czech Academy of Sciences and by the funds from Project CoGraDS
- CZ.02.1.01/0.0/0.0/15\_003/0000437 from the European Structural
and Investment Fund and the Czech Ministry of Education, Youth and
Sports (M\v{S}MT). 

\bibliographystyle{utphys}
\addcontentsline{toc}{section}{\refname}\bibliography{HiggsDust}

\end{document}